\documentclass[aps,prb]{revtex4}
\usepackage{graphicx}
\newcommand{\beq}{\begin{equation}}
\newcommand{\eeq}{\end{equation}}
\begin{document}
\draft

\title{BCS pairing in Fermi systems with several flavors}
\author{C. Honerkamp$^{(1)}$ and W. Hofstetter$^{(2)}$}
\address{$^{(1)}$ Max Planck Institute for Solid State Research, D-70569 Stuttgart, Germany
\\ $^{(2)}$ Department of Physics, Massachusetts Institute of Technology, Cambridge MA 02139, USA}  
\date{\today}
\begin{abstract} 
Motivated by the prospect of Bardeen-Cooper-Schrieffer (BCS) pairing in cold fermionic gases we
analyze the superfluid phase of 3 fermionic flavors in the attractive
Hubbard model.  We show that there are several low--lying collective
pairing modes and investigate their damping due to the partially
gapless nature of the single-particle spectrum. Furthermore we
analyze how these modes show up in the density response of the
system. Apart from the Anderson-Bogoliubov phase mode of the pairing
between two flavors, the dynamical structure factor contains
signatures of the gapless third flavor. This picture is found to be
robust against perturbations that break the global SU(3)-symmetry of
the Hamiltonian.

\end{abstract}
\maketitle

\section{Introduction}
The experimental realization of Bose--Einstein condensation \cite{bec} 
has opened up a new avenue of research in systems of ultracold atoms. 
Recently, improved cooling techniques have also led to 
degenerate Fermi clouds \cite{deMarco,truscott,fer3,fer4}. 
An active search is now under way to implement a BCS transition 
of degenerate fermionic gases to a superfluid state \cite{stoof}, and very
recently first indications of pairing of $^{40}$K atoms have been 
reported\cite{regal03,zwierlein,regal04}. 
Although interatomic interactions are generally weak, it has been 
become possible to approach \emph{strongly correlated} phenomena  
either via Feshbach resonances \cite{inouye,timm,holland,regal03} or optical lattices \cite{boseth}. 
In this work we focus on the latter possibility. 
In a pathbreaking recent experiment, a 3d optical lattice has been used to  
experimentally realize the transition between a bosonic superfluid 
and a Mott insulator\cite{greiner}.
In combination with fermionic atoms, optical lattices offer 
the intriguing perspective of studying solid--state phenomena 
like high--temperature superconductivity in a new context 
\cite{hofstetter}. 

Compared to electrons in solids, 
atomic systems offer new internal 
degrees of freedom that can lead to states 
of matter which do not have obvious counterparts in the physics 
of interacting electrons. 
For alkali atoms, nuclear spin $I$ and electron spin $S$ are combined 
in a hyperfine state with total angular momentum $F$. 
While typical electronic systems are constrained to SU(2) 
spin rotational symmetry, the atomic total angular momentum $F$ 
can be larger than 1/2, resulting in $2F+1$ hyperfine states 
differing by their azimuthal quantum number $m_F$. 
In a magnetic trap, only a subset of these 
$2F+1$ states (the \emph{low--field seekers}) can be trapped, 
but this constraint can be avoided using all--optical traps \cite{granade}. 
In fact, coexistence of the three hyperfine states 
$|F=9/2, m_{F}=-5/2,-7/2,-9/2\rangle$ of $^{40}K$ in an optical trap 
has already been demonstrated, 
with tunable interactions due to Feshbach resonances between 
$m_{F}=-5/2/-9/2$ and $m_{F}=-7/2/-9/2$, respectively \cite{regal03}.
A situation with strong attractive interaction between all three 
components can be realized e.g. for the spin polarized states with 
$m_{s}=1/2$ 
in $^{6}$Li where the triplet scattering length $a=-2160a_{0}$ 
is anomalously large \cite{abraham}. 

Optical lattices are created by a standing light wave leading
to a periodic potential for the atomic motion of the form
$V(x)=V_{0} \sum_{i} \cos^{2}(k x_{i})$
where $k$ is the wavevector of the laser, $i$ labels the spatial
coordinates and the lattice depth $V_{0}$ is usually measured in units of the 
atomic recoil energy $E_{R}=\hbar^{2} k^{2}/2m$. 
In the following we will consider the 2D case where
$i=1,2$. It has been shown \cite{boseth} that
the \emph{Hubbard model} with a local density--density interaction
provides an excellent description of the low--energy physics.
Here we are interested in a situation where fermionic atoms with
$N$ different hyperfine states (further on denoted as ``flavors'') $m$ are loaded into the
optical lattice.
We thus consider a Hubbard Hamiltonian
\begin{equation} 
H =  -t \sum_{m,\langle ij \rangle} 
    \left[c^\dagger_{i,m} c_{j,m} + c^\dagger_{j,m} c_{i,m} \right] + 
\frac{U}{2} \sum_i n_i^2 \, .  \label{hubb}
\end{equation} 
Here $n_i=\sum_{m} n_{i,m}$ is the total number
density of atoms on site $i$ which can be written in terms of the fermionic 
creation and annihilation operators,  
$n_{i,m} = c^\dagger_{i,m} c_{i,m}$.  
The interaction (second term in Eq. \ref{hubb}) 
is invariant under local U($N$) rotations of the $N$
flavors with different $m$.  The hopping term of the atoms between 
nearest neighbors $\langle ij \rangle$
reduces the invariance of 
$H$ to a global U($N$) symmetry. Stripping off
the overall U(1) phase factor, we arrive at the SU($N$) Hubbard model. 
In the optical lattice  the Hubbard parameters are 
$t=E_{R} \left(2/\sqrt{\pi}\right) \xi^{3} \exp\left(-2 \xi^{2}\right)$ 
and $U=E_{R} a_{s} k \sqrt{8/\pi} \xi^{3}$ where 
$\xi=\left(V_{0}/E_{R}\right)^{1/4}$. $a_{s}$ is the $s$--wave 
atomic scattering length. 

The fermionic SU($N$) Hubbard model on the two-dimensional (2D) 
square lattice was studied for repulsive interactions $U>0$ 
in the large-$N$ limit \cite{marston} in the early days of 
high--$T_c$ 
superconductivity, mainly as a controllable limit connected to the 
then 
physically relevant case $N=2$. 
Affleck et al.~\cite{affleck} already discussed realizations of SU(4) 
using the nuclear spin of $^{21}$Ne.
A generalized SU($N$) model could describe 
orbitally degenerate electronic states in crystals, but it is likely 
that 
different overlaps between the orbitals 
pointing in distinct lattice directions will break the SU($N$) 
invariance.   In a previous work\cite{hoho} we have shown that exotic states like staggered flux states and partially incompressible flavor density wave states may occur in the repulsive 2D Hubbard model for finite $N>2$. We also pointed out that in the attractive case for $N>2$, the onsite Cooper pairing breaks more symmetries than just a global U(1) symmetry. This gives rise to new collective modes in the paired state.

In this paper we focus on the attractive regime $U<0$ and the question 
how the spin--1/2 BCS state will be generalized for a number 
of flavors $N>2$. 
We shall assume weak to intermediate local attractions  such that a treatment within BCS theory is qualitatively valid.   
We focus mostly on $N=3$ where we analyze the ground state manifold 
using mean--field theory and determine the collective modes by a generalized random phase approximation (RPA). 
We show that the density response obtained by \emph{Bragg scattering}  
provides a clear experimental signature for this new superfluid ground state.
Although our considerations take place in the framework of the Hubbard lattice model, the qualitative features of the excitation spectrum should remain the same for continuum systems with attractive interactions between three flavors. 

\begin{figure}
\includegraphics[width=.6\textwidth]{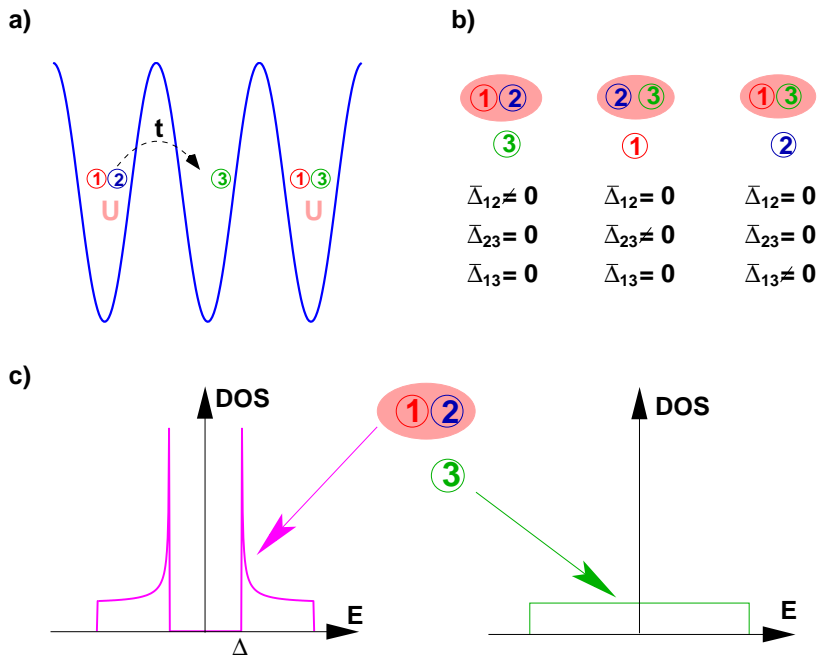}
\caption{a) Hubbard model for three flavors of fermions. The particles can hop with amplitude $t$ between neighboring sites of the optical lattice and interact on the same lattice site with an onsite interaction $U$. b) Three possible types of onsite pairings for three flavors of fermions with attractive interaction. In the mean-field theory for the  SU(3) symmetric case all three types of pairings and superposition thereof have the same  ground state energy. c) Density of states (DOS) for the three fermionic excitation branches of the SU(3)-paired state (schematic, lattice effects neglected). In the gauge with only $\bar\Delta_{12}\not=0$, the unpaired flavor 3 is gapless while the two branches involving flavors 1 and 2 see an energy gap $\Delta$.} 
\label{su3}
\end{figure} 

\section{Mean-Field Theory}
Let us briefly describe the results of BCS mean field theory for the SU($N$)-model with attractive contact (Hubbard) interaction.
Within the mean-field approximation these results are quite general and hold independent of the spatial dimension, also for continuum systems without underlying lattice. Parts of the analysis are already contained in a brief form in Ref. \onlinecite{hoho}.

The Hamiltonian we study is 
\begin{equation} 
H = \sum_{\vec{k},\alpha} (\epsilon_{\vec{k}} - \mu_\alpha) 
c^\dagger_{\vec{k},\alpha} c_{\vec{k},\alpha} +
\frac{U}{2V} \sum_{\vec{k},\vec{k}',\vec{q}\atop \alpha \beta} c^\dagger_{\vec{k}+\vec{q},\alpha} c^\dagger_{-\vec{k},\beta} c_{-\vec{k}',\beta}  c_{\vec{k}+\vec{q},\alpha} \, .  
\label{hubbham}\end{equation} 
with a kinetic energy $\epsilon_{\vec{k}} =\epsilon_{-\vec{k}}$, 
flavor indices $\alpha$, $\beta \in 1\dots N$ and chemical potentials $\mu_\alpha$.
$V$ denotes the number of lattice sites. We introduce a pairing mean-field
\[ \Delta_{\alpha \beta} = -\frac{U}{V} \sum_{\vec{k}} 
\langle  c_{\vec{k}\alpha}   c_{-\vec{k}\beta} \rangle   \] and consider even-parity pairing $\Delta_{\alpha \beta} = - \Delta_{\beta 
\alpha} $ (see also Fig. \ref{su3}). Non-trivial odd parity solutions are not possible for local attractions. 
The mean--field Hamiltonian then becomes
\[  H_{U,\mathrm{mf.}}= -\frac{1}{2} \sum_{\vec{k}, \alpha ,\beta}  
c^\dagger_{\vec{k}\alpha}   c^\dagger_{-\vec{k}\beta} \Delta_{\beta 
\alpha} + h.c.  \]
For $N>2$ the even parity gap functions 
$\Delta_{\alpha \beta}$ transform non-trivially under SU($N$). 
This means that depending on the global gauge of the fermions, $\Delta_{\alpha \beta} $ takes different values. This is different from the SU(2) case. There even parity gap functions are singlets and invariant under spin rotations and transform with the trivial representation. 
From a particle physics perspective we can compare the BCS pairing in the SU(3) case with the pairing of two quarks or two antiquarks. The quarks transform according to the color-SU(3) which is a local gauge symmetry in this case. 
Bound states  of two quarks or two antiquarks carry color-SU(3) charge and do apparently not exist in nature. In contrast with that pairs of quarks and anitquarks, e.g. pions, can be color-SU(3) singlets.

Without much group theoretical input we can read off the dimension of the representation in the SU($N$) case: an antisymmetric complex $N\times N$ matrix $ \Delta_{\alpha \beta}$ has $N(N-1)/2$ independent complex components. For example for $N=3$, $ \Delta_{\alpha \beta}$ transforms according to a 3D representation. This is consistent with the decomposition of the product $3 \otimes 3 =\bar{3}\oplus 6$. Here 3 denotes the irreducible representation under which the flavor spinor $(c_{\vec{k},1}, c_{\vec{k},2},  c_{\vec{k},3})$ is transformed, and $\bar{3}$ is the complex conjugate representation of $3$. This can be checked explicitely when we write (dropping the wavevector indices for a moment)
\[ \Delta_\alpha = {1 \over 2} \, \epsilon_{\alpha \beta \gamma} \langle c_\beta c_\gamma \rangle  = \left( \begin{array}{r} \Delta_{23} \\ - \Delta_{13} \\ \Delta_{12} \end{array} \right)  \]
and analyze the transformation $\tilde{\Delta}_\alpha = D(U)_{\alpha \beta} \Delta_\beta$ when the fermions are transformed as $\tilde{c}_\alpha = U_{\alpha \beta} c_\beta$, where $U$ is an SU(3) matrix. Expanding $U_{\alpha \beta}= \delta_{\alpha \beta} + i u_{\alpha \beta}$ and $D_{\alpha \beta}= \delta_{\alpha \beta} + i d_{\alpha \beta}$ for an infinitesimal transformation and using the traceless hermitian nature of $ u_{\alpha \beta}$ and $d_{\alpha \beta}$, we obtain $d_{\alpha \beta} =  - u_{\alpha \beta}^*$, i.e. $D=U^*$. This transformation property implies that any 3-vector $\Delta_\alpha$ with $\sum_\alpha |\Delta_\alpha|^2 = \Delta_0^2 $ can by a global gauge change be mapped onto $(0,0,\Delta_0)$, where $\Delta_{12}=\Delta_0$ and $\Delta_{13}=\Delta_{23}=0$. 
In this gauge the mean-field solution is simply the solution for the SU(2) BCS problem in flavors 1 and 2 plus the third flavor that remains unpaired and thus ungapped. Therefore in the ground state and weak coupling $N_F |U| \ll 1$
\[ \Delta_0 =  W \, e^{-1/(N_F |U|)} \, , \]
where $N_F$ is the density of states at the Fermi level and $W$ the bandwidth within which the attraction $U<0$ is active.  For $3/8$ band filling and $U=-4t$, the mean-field $T_c$ is $\sim 0.17t$.
The quasiparticle spectrum contains three branches, 
\[ E_{1,2}= \pm \sqrt{\epsilon_{\vec{k}}^2 + \Delta_0^2} \qquad \mbox{and} \qquad  E_{3} = \epsilon_{\vec{k}}\, , \]
 and is independent of the global gauge. Therefore the paired mean-field ground state of the SU(3) BCS problem has a full Fermi surface and only two thirds of the excitations are gapped (see also Fig. \ref{su3}).

For SU(2), the ground state is degenerate with respect to the 
global phase of the gap function, and long-wavelength variations of 
the latter are gapless in absence of long-range forces. In the 
SU($N$) case we find a higher degeneracy of the ground state 
and more gapless modes as more generators are broken.
As argued above, for SU(3) all gap functions    
with  the same $\Delta_0^2= \sum_{\alpha \beta} |\Delta_{\alpha 
\beta}|^2$ are degenerate and have the same total density of states.
Apart from the global phase there are four additional gapless 
modes, 
two associated with the internal phases between $\Delta_{12}$, 
$\Delta_{13}$ and $\Delta_{23}$, and two modes modulating 
$|\Delta_{12}|$, 
$|\Delta_{13}|$ and $|\Delta_{23}|$ with fixed $\Delta_0$.

The gauge with only $\Delta_{12}\not= 0$ and $\Delta_{23}=\Delta_{13}=0$ 
makes the symmetry breaking  pattern obvious. 
The original symmetry group of the problem SU(3) 
$\otimes$ U(1) has nine generators. This gets broken down to an SU(2) 
symmetry in flavor 1 and 2, leaving $\Delta_{12}$ invariant, and an additional 
U(1) that acts on the phase of the unpaired flavor 3. This leaves 5 
generators broken, yielding the collective modes described above. 

The coexistence of a full Fermi surface with a superconductor should 
have interesting consequences. Some of these will be analyzed in the following sections.

Let us briefly comment on perturbations that break the SU(3) symmetry of the Hamiltonian. 
A weak magnetic field will cause a Zeeman splitting of the hyperfine levels with 
different $m_F$. This is equivalent to having different chemical potentials 
for the three flavors. As the flavor densities become unequal, 
the mean-field solution for zero total momentum pairing 
retains a nonzero $\Delta_0$ if the Zeeman splitting between two neighboring hyperfine 
levels $m_F$, $m_F+1$ is smaller than $\Delta_0$. 
We do not consider the possibility of pairing with non-zero total momentum, leading 
to a so-called Larkin-Ovchinikov-Fulde-Ferrell (LOFF) state\cite{loff}. For the continuum model 
it has been found\cite{liu} that zero-total-momentum pairing gives the lower energy 
for sufficiently strong attraction.  
If the interaction remains the same between all flavors, the zero-total-momentum pairing 
will choose those hyperfine levels that have the largest density of states $\rho_\alpha$ 
near the respective Fermi levels, i.e. if $\rho_1 > \rho_2 > \rho_3$, the pairing amplitudes 
pick the solution $|\Delta_{12}|= \Delta_0$ and $\Delta_{23}=\Delta_{13}=0$. In the case with 
unequal strengths of the attraction between the flavors, the pairing amplitude chooses the 
strongest interaction. This means for $-U_{12} > -U_{23}$ and $-U_{12} > -U_{13}$, we again 
have $|\Delta_{12}|= \Delta_0$ and $\Delta_{23}=\Delta_{13}=0$. 

For SU(4), the  antisymetric matrix $\Delta_{\alpha \beta}$ has 6 independent components 
$\Delta_i$, $i=1,\dots,\,  6$,  which transform under a 6-dimensional representation. 
Then with an SU(4) rotation we cannot always map a general 6-vector $\Delta_i$ onto a 
vector $\propto \Delta_0 \cdot (0,0,0,0,0,1)$. Correspondingly the degeneracy 
of the ground state is subject to more constraints than just constant 
$\Delta_0$. The mean-field solutions have 
$|\Delta_{12}|=|\Delta_{34}|$, 
$|\Delta_{13}|=|\Delta_{24}|$ and  $|\Delta_{14}|=|\Delta_{23}|$. 
The single particle spectrum is fully gapped. 

\section{Generalized RPA for the superfluid state\label{RPA}}
Here we describe a generalized random phase approximation (RPA) scheme that allows us to analyze pairing and density fluctuations in the attractive SU(3) Hubbard model. 

The theoretical description of the interference of different collective fluctuations of a many-fermion system is a complex problem. 
In principle the interacting two-particle scattering vertex contains all the information about possible collective modes. 
However, for general situations where charge, spin and pairing channels are
important, the calculation of this object, e.g.~using a Bethe-Salpeter
equation, is rather involved even if self-energy effects are neglected,
because both particle-particle and particle-hole diagrams have to be taken into account.   
Fortunately, in many situations one can with good confidence restrict the
analysis on certain channels and apply a Hubbard-Stratonovitch (HS) decoupling
of the interaction in physically reasonable way. In order to motivate our
approach let us sketch our expectations regarding the collective excitations in the SU(3) case with local attractive interactions.

The pairing channel contains soft collective modes due to the spontaneous symmetry breaking and we expect that the density of the fermions will couple to these modes. For example for superconductors with long range Coulomb interactions it is known\cite{anbog} that the density response ``eats'' the Goldstone phase mode 
of the superconductor by pushing it up to the plasma frequency, which is the collective mode of 
the charge density. 
In our case of an attractive Hubbard interaction there is no long range force, but the coupling of the phase mode to the partially gapless density fluctuations could still lead to observable effects. 
In particular one may wonder whether the phase mode gets strongly damped by the third flavor. 
A similar obliteration of a collective mode occurs in the normal state for $U<0$. There in RPA the zero sound mode of the three flavors is located inside the particle-hole continuum and thus it is strongly Landau damped. 
In the case of three flavors with pairing of two flavors, the ungapped flavor could still maintain a zero sound mode that sharpens below the transition as a part the quasiparticles become gapped and the damping is reduced. 

In the flavor channel, in analogy with the spin channel of the SU(2) case, the sign of the 
effective interaction is reversed and for $U<0$ there are two flavor modes with linear dispersion 
above the particle-hole continuum. In the normal state  these modes can be understood as relative 
density 
oscillations $\propto \rho_1+\rho_2 -2 \rho_3$  and $\propto   \rho_1 - \rho_2$ of the
three flavors that keep the total density $\rho_1+\rho_2+\rho_3$ fixed. Hence
these flavor modes are invisible in the density response. 
In the superfluid state with only $\Delta_{12} \not= 0$ however, the density response of the gapped flavors 1 and 2 is different from that of the gapless flavor 3, and the mode  $\propto \rho_1+\rho_2 -2 \rho_3$ couples to the total density. This could make the flavor mode observable in the density response.

Summarizing this short discussion, we state that a Hubbard-Stratonovitch decoupling scheme should include the pairing degrees of freedom and the individual densities of the three flavors.
Let us now describe a formalism that allows a treatment of the issues raised above. It is a straightforward generalization of the methods described e.g. in Ref. \onlinecite{nagaosa}. We start with the attractive Hubbard Hamiltonian of Eq. \ref{hubbham}. 
Next we promote the fermionic operators to Grassmann numbers and introduce the combined index $k=(i \omega , \vec{k})$. Then the partition function for the system is given by
\begin{equation}
Z = \int DcD\bar{c} \; \exp \left[ - \sum_{k,\alpha} \bar{c}_{k\alpha} 
(i \omega + \epsilon_{\vec{k},\alpha} - \mu_{\alpha}) \, c_{k\alpha} - 
\frac{U}{2V} \sum_{k,k',q \atop \alpha \beta} \bar{c}_{k+q,\alpha} \bar{c}_{-k,\beta} 
c_{-k',\beta}  c_{k'+q,\alpha} \right] \, . 
\end{equation}
Now we decouple the local attraction in pairing, density and flavor channels. We write
\begin{equation} 
H_U = \frac{U}{2} \sum_{i \atop \alpha \beta} n_{\alpha,i} n_{\beta,i}  = \frac{U_p}{2} \sum_{i \atop \alpha \beta}  n_{\alpha ,i} n_{\beta ,i} + \frac{U_\rho}{2} \sum_{i \atop \alpha \beta} n_{\alpha ,i} n_{\beta ,i} \, . \label{decomp}
\end{equation}
Here the first term is decoupled in the onsite pairing channel with a coupling constant $U_p<0$, while the second term with $U_\rho = U- U_p < 0$ is decoupled with local Hubbard-Stratonovitch fields coupling to the densities of the three flavors. To this end we write in the pairing channel
\begin{eqnarray} 
\exp \left( \frac{|U_p|}{2} \sum_{i \atop \alpha \beta}  n_{\alpha ,i} n_{\beta ,i} \right)& =&
\int D\tilde\Delta_iD\tilde\Delta^*_i \exp \left\{  \frac{1}{2} \sum_{i \atop \alpha \beta} 
\left[ \bar{c}_{i,\alpha} \bar{c}_{i,\beta} \tilde\Delta_{\alpha\beta,i}
+ \tilde\Delta^*_{\alpha \beta,i} c_{i,\beta}  c_{i,\alpha} \right] - \frac{1}{2|U|} \sum_{i \atop \alpha \beta} | \Delta_{\alpha \beta,i} |^2 \right\}
\\ &=&
\int D\tilde\Delta_iD\tilde\Delta^*_i \exp \left[ -S_\Delta (c,\bar c,\Delta, \Delta^*)\right]
\end{eqnarray} 
Searching the saddle point with respect to the Hubbard-Stratonovitch field $\tilde\Delta_{\alpha \beta}$ reproduces mean-field theory. Let us write
\begin{equation} \tilde\Delta_{\alpha \beta,i} = \bar\Delta_{\alpha \beta}  + \Delta_{\alpha \beta,i}  \end{equation}
where $\bar\Delta_{\alpha \beta} $ denotes the static site-independent saddle point solution and $\Delta_{\alpha \beta,i}$ are the fluctuations around it. The saddle point solution $\bar\Delta_{\alpha \beta}$ can be absorbed into the mean-field action $S_{\mathrm{MF}}$.
Next we split up $\Delta_{\alpha \beta}$ in real and imaginary parts, $\Delta_{\alpha \beta,i}= \Delta^r_{\alpha \beta,i} + i \Delta^i_{\alpha \beta,i}$. Going over to wavevector-frequency space we obtain the pair fluctuation action
\begin{eqnarray} 
S_\Delta (c,\bar c,\Delta^r, \Delta^i) &=&
- \frac{1}{2V} \sum_{k,q \atop \alpha \beta} 
\left[ \bar{c}_{k+q,\alpha} \bar{c}_{-k,\beta} (\Delta^r_{\alpha\beta}(q) + i \Delta^i_{\alpha\beta}(q) )
+ ( \Delta^{r*}_{\alpha \beta}(q)- i \Delta^{i*}_{\alpha \beta}(q))  c_{-k,\beta}  c_{k+q,\alpha} \right] \nonumber \\ && + \frac{1}{2|U|} \sum_{q \atop \alpha \beta} \left[ | \Delta^r_{\alpha \beta} (q) |^2 + | \Delta^i_{\alpha \beta} (q) |^2 \right]
 \, . \label{mfaction} \end{eqnarray}
In the flavor and density channel we write the interaction as a sum of squares,
\[ \frac{1}{2} \sum_{i \atop \alpha \beta} n_{\alpha ,i} n_{\beta ,i} = \left( \frac{n_{1,i}+n_{2,i}+n_{3,i}}{\sqrt{3}} \right)^2 - \left( \frac{n_{1,i}+n_{2,i}-2n_{3,i}}{\sqrt{12}} \right)^2 - \left( \frac{n_{1,i}-n_{2,i}}{\sqrt{4}} \right)^2 \, , 
\]
and use the HS identity
\begin{eqnarray} && \exp \left\{  |U_\rho| \left[ \left( \frac{n_{1,i}+n_{2,i}+n_{3,i}}{\sqrt{3}} \right)^2 - \left( \frac{n_{1,i}+n_{2,i}-2n_{3,i}}{\sqrt{12}} \right)^2 - \left( \frac{n_{1,i}-n_{2,i}}{\sqrt{4}} \right)^2 \right] \right\} \nonumber \\ 
&& \hspace{1cm} = \int D\rho_{t,i} \, D\rho_{123,i} \, D\rho_{12,i} \, \exp 
\left\{  - \frac{1}{4|U_\rho|} \left[ \rho_{t,i}^2 +\rho_{123,i}^2 +\rho_{12,i}^2 \right] \right.  \nonumber \\ 
&&  \hspace{3cm} +  \left. \rho_{t,i} \frac{n_{1,i}+n_{2,i}+n_{3,i}}{\sqrt{3}}  +i \rho_{123,i} \frac{n_{1,i}+n_{2,i}-2n_{3,i}}{\sqrt{12}} +i\rho_{12,i} \frac{n_{1,i}-n_{2,i}}{\sqrt{4}} \right\} \nonumber \\ && \hspace{1cm} =
\int D\rho_{t,i} \, D\rho_{123,i} \, D\rho_{12,i} \, \exp \left[ -S_\rho (c,\bar c,\rho_t, \rho_{12}, \rho_{123}) \right] \label{srho}
\end{eqnarray}
with the real site-dependent HS fields $\rho_{t,i}$, $\rho_{123,i}$ and $\rho_{12,i}$.
Here $\rho_{t,i}$ couples to the total density, and $\rho_{12,i}$ and $\rho_{123}$ couple to out-of-phase oscillations of the flavors. We will denote these as flavor modes. For all situations we study, the static components of the flavor fields and also the density field $\rho_t (\vec{q},\omega=0)$ away from $\vec{q} \not= 0$ remain zero, i.e. we do not study flavor or charge density wave states. In the case where $U_\rho \to 0$, the internal $\rho_a$ fluctuations get pinned at zero and do not contribute to the partition function. 
In Fourier space we have 
\begin{eqnarray}
S_\rho (c,\bar c,\rho_t, \rho_{12}, \rho_{123}) =
-  \sum_{q,a} \rho_a (q) N_a (-q) + \frac{1}{4|U_\rho|} \sum_{q,a}  |\rho_a (q)|^2
\end{eqnarray}
where the index $a$ labels the channels $t$, $12$ and $123$ and 
$N_t(q)= \frac{1}{V} \sum_{k,\alpha} \bar{c}_{k+q,\alpha} c_{k,\alpha}$, 
$N_{12}(q)= \frac{1}{2 V} \sum_{k,\alpha} \left( \bar{c}_{k+q,1} c_{k,1}- \bar{c}_{k+q,2} c_{k,2} \right) $    and $N_{123}(q)= \frac{1}{\sqrt{12} V} \sum_{k,\alpha} \left( \bar{c}_{k+q,1} c_{k,1}+ 
\bar{c}_{k+q,2} c_{k,2} - 2\bar{c}_{k+q,3} c_{k,3} \right) $. 
In the functional integrals over the HS fields $\Delta^{r/i}_{\alpha \beta}$ and $\rho_a$ we have 
to keep in mind that their real space versions are real fields. Therefore we have  
$\Delta^{r/i,*}_{\alpha \beta}(q)= \Delta^{r/i}_{\alpha \beta}(-q) $  and $\rho_a^* (q) = \rho_a (-q)$. 

Finally we  add an external total density field $\rho(q)$. It couples to the total density of the fermions in the same way as $\rho_t$. However as external field it is not integrated over in the partition function and there is no contribution $\propto 1/U_\rho$. The action for this coupling is 
\begin{eqnarray}
S_{\mathrm{ext.}} (c,\bar c) = -  \frac{1}{V} \sum_{k,q \atop \alpha} \rho (-q) \, \bar{c}_{k+q,\alpha} c_{k,\alpha}   \, . \label{rhoext}
\end{eqnarray}

The quadratic fermionic parts of these four contributions to the action, $S_{\mathrm{MF}}$, $S_\Delta$, $S_\rho$ and $S_{\mathrm{ext.}}$ can be cast into a $6\times 6$ Nambu matrix form using a 6-vector $C^{k}_\alpha = ( c_{k,1},c_{k,2},c_{k,3}, \bar{c}_{-k 1},\bar{c}_{-k 2},\bar{c}_{-k 3})$.  This way we have doubled the fermionic degrees of freedom and have to restrict the $\vec{k}$-space sums to half of the Brillouin zone.
Up to the non-fermionic parts the action can be  written as (using the summation convention 
in wavevectors $k$, the Nambu-flavor index $\alpha$ and the fluctuation index $A$)
\begin{equation} S^c_F = \bar{C}^k_\alpha (G^{-1})_{\alpha \beta}^{kk'} C^{k'}_\beta
- \bar{C}^k_\alpha F^{kk'}_A\ ^Af_{\alpha \beta}^{kk'} C^{k'}_\beta \end{equation}
Here $G^{-1}$ is the inverse Green's function of the saddle point theory, containing $\bar\Delta_{\alpha\beta}$. $^Af_{\alpha \beta}^{kk'}$ is the vertex factor coupling  a flavor or pairing HS field $F_A$ with wavevector/frequency $k-k'=q$ to the fermion bilinear. As we have 6 pairing fields $\Delta^{r/i}_{\alpha\beta}$, 3 flavor/density HS fields and 1 external source field, 
$A=1,\dots ,10$.
Integrating out the fermions yields the determinant of this quadratic kernel, or its logarithm if we put the result back into the exponent. Then we use $\log \det S^c_F = \mbox{tr} \log S^c_F $ and obtain the following contribution to the action for the fluctuations
\begin{equation}
S^c_F = - \mbox{tr} \log \left[  (G^{-1})_{\alpha \beta}^{kk'} -  F^{kk'}_A\  ^Af_{\alpha \beta}^{kk'} \right] \end{equation}
Our goal is to expand this expression to second order in $\Delta_{\alpha \beta}^{kk'} $. 
Together with the contributions from the last term in the action (\ref{mfaction}) this will give a quadratic action for the fluctuations.  The terms linear in the fluctuations cancel provided we expand around the saddle point.
Next we write $G^{-1} -  F_A \ ^Af = G^{-1}(1-  G F_A \, ^Af)$ and use $\log (1-x) = -x - \frac{x^2}{2} - \dots $. Thus in order to extract the fermionic contribution to the fluctuation action, we  have to evaluate 
\begin{equation}
S_F^{(2),c} = \frac{1}{2}  \,  G^{k}_{\alpha\beta''} F^q_A  \  ^Af^{q}_{\beta''\beta'} G^{k-q}_{\beta'\beta'''} F^{-q}_{A'} \ ^{A'}f^{-q}_{\beta'''\alpha} 
\end{equation}
Here we have used the fact that $G^{k,k'}_{\alpha\beta'}$ is diagonal in $k,k'$ and denoted $k-k'=q$.
Diagrammatically, these contributions correspond to one-loop bubbles made out of two 
Green's functions and perturbations $F^q_{A}$  
at the vertices. Including the non-fermionic contributions the quadratic fluctuation action reads
\begin{equation}
S_F^{(2)}=   F^q_{A}  \left[ d_A \delta_{A A'}   
+ \frac{1}{2}\ ^{A}f^{q}_{\beta'''\alpha } G^k_{\alpha \beta''} G^{k+q}_{\beta' \beta'''}  \ ^{A'}f^{-q}_{\beta''\beta'} \right]         F^{-q}_{A'}  
\label{s2f}
\end{equation}
Here the diagonal factors $d_A$ are $d_A= V/|U_p|$ for the pairing HS fields and  $d_A= V/4|U_\rho|$ for the flavor-density HS fields. For the external source field $\rho(q)$, $d_{A=10} =0$, and the only contributions are due to the fermionic bubbles. 
It is convenient to group the HS fields into a 10-vector
\begin{equation} 
\hat{F} (q) =
\left( \begin{array}{c} 
\Delta^r_{12}(q) \\  \Delta^i_{12}(q) \\ 
\Delta^r_{13}(q) \\   \Delta^i_{13} (q)  \\
\Delta^r_{23}(q) \\   \Delta^i_{23} (q)  \\ 
\rho_t(q) \\ \rho_{123} (q) \\ \rho_{12} (q) \\ \rho (q) \end{array} \right)
\end{equation}
With this Eq. \ref{s2f} reads
\begin{equation}
S_F^{(2)} = \sum_q \hat{F} (q) \hat{M} (q) \hat{F} (-q)  \label{meq}
\end{equation}
This quadratic action couples the internal degrees of freedom among each other and to the external field. Like in any other RPA, only modes with the same frequency/wavevector $q$ interact with each other.
If we are interested in the Green's function of a specific fluctuation $F$ like the amplitude or phase of a $\Delta_{\alpha\beta}$ in absence of the external field, we integrate out the remaining 8 internal degrees of freedom yielding
\begin{equation}
 M^{r}_{FF} = M_{FF}  - \sum_{F'=1, \dots 9 \, \mathrm{w.o.} F\atop F''=1, \dots 9 \, \mathrm{w.o.} F} M_{FF'}M^{-1}_{F'F''} M_{F''F} \, , \label{MFren}
\end{equation}
Here the sums $F'$ and $F''$ run over the other 8 internal degrees of freedom, excluding $F$ and the external source $\rho$. The Green's function for fluctuation $F$ is given by the inverse of $M^{r}_{FF}(q)$, and the corresponding spectral function is obtained after analytical continuation $\nu \to \omega +i \delta$ as $A_F (\vec{q},\omega) = \mbox{Im} \, \left[ M^{r}_{FF}(\vec{q},\omega)\right]^{-1}$. 

Regarding experimental manifestation of the collective excitations we are interested in the total density response. This is given by the renormalized quadratic part in $\rho(q)$ after integrating out the 9 internal HS fluctuation fields. Doing the Gaussian integral yields
\begin{equation}
 M^{r}_{\rho \rho} (q) = M_{\rho \rho} (q) -   \sum_{F=1, \dots 9 \atop F'=1, \dots 9} M_{\rho F'}M^{-1}_{F'F''} M_{F''\rho} \, , \label{total_density}
\end{equation}
where this time the indices $F'$ and $F''$ go through the 9 internal degrees of freedom.  The total density susceptibility $\chi_\rho(q)$ is then given by the double derivative of the remaining exponential with respect to the source fields $\rho(q)$ and $\rho(-q)$ and the sources set to zero afterwards. 
This simply yields  
\[ \chi_\rho(q) = M^r_{\rho\rho}(q) \, . \]
The imaginary part is the total density spectral function Im $\chi_\rho (\vec{q}, \omega)$.

The density susceptibility $\chi_\rho (\vec{q}, \omega)$ is relevant for experiments in ultracold atoms, where the dynamical structure factor 
\beq
S(q,\omega) = \sum_n |\left(\rho_q^\dagger\right)_{n0}|^2 \delta(\omega - \omega_{n0})  
\eeq
can be measured via \emph{Bragg scattering} \cite{stenger,ketterle}. 
Here $\omega_{n0} = E_n - E_0$ denotes the many--body excitation energies relative to the ground state. 
Within this technique, two laser beams with wavevector-- and frequency--difference $k$ and $\omega$ 
are used to create a time-dependent potential (\emph{light grating}) 
\beq
V_{\rm mod}(x,t) = V \cos(q r - \omega t) 
\eeq 
leading to scattering of atoms due to two--photon processes. 
The resulting transition rate per atom is given by 
\beq
W/N = \frac{\pi}{2} V^2 S(q,\omega). 
\eeq
At $T=0$ the structure factor is related to the density susceptibility according to 
\beq
\mbox{Im}\, \chi_\rho (q,\omega) = -\pi \left\{S(q,\omega) - S(q, -\omega)\right\}
\eeq
i.~e.~at positive frequencies we have $S(q,\omega) = -\frac{1}{\pi} \mbox{Im}  \chi_\rho (q,\omega)$.  
From our calculation of the density response in Eq.~(\ref{total_density}) we 
can thus extract the Bragg scattering rate. 

In the decomposition (\ref{decomp}) we will later choose $U_p=U_\rho=U/2$, but in principle this 
decomposition is arbitrary and all choices are equally exact as long as the full action is treated.  
Yet within the Gaussian approximation we have to resort to in the further evaluation of 
the partition function, different decompositions can in principle lead to different 
saddle point solutions. This is a well known ambiguity of the Hubbard-Stratonovitch formalism. 
In our case however the qualitative results do not depend on the precise partitioning of 
the interaction. Sufficiently away from half band filling, the ground state in the 
saddle point approximation will always be given by the BCS mean-field solution, provided $U_p<0$. 
The gap value  $\Delta_0$ will be reduced corresponding to a weaker pairing 
interaction $U_p = U - U_\rho $.

\section{Collective modes from pairing fluctuations}
To begin with, let us set $U_\rho=0$ and decouple the interaction in the pairing channel only. This allows us to analyze the pairing fluctuation as internal degrees of freedom without interaction to other collective fluctuations. 
The coupling to the flavor densities and the total density response will then be analyzed in the following section. Unless stated otherwise, all numerical results shown in the following section were obtained for the parameters $U_p=-4t$ and $\langle n \rangle =3 \cdot 0.185 = 0.556$ particles per site. The mean-field critical temperature for this case is $T_c \approx 0.07t$ and $\Delta_0 \approx 0.1t$ at zero temperature. We analyze the spectral functions of the fluctuations for wavevectors $\vec{q}=(q,0)$ with small $q$ along the $x$-axis. 
As the Fermi surface is anisotropic, different directions in $\vec{q}$-space
give slightly different results. However we have checked that the qualitative
features we show in the following 
are common to all cuts through the Brillouin zone. 
\subsection{SU(3)-symmetric case}
First we consider the SU(3) symmetric case. As explained in Sec. II, the mean-field ground state is degenerate for all choices of saddle points $\bar\Delta_{\alpha\beta}$ with  ${1 \over 2} \sum_{\alpha,\beta} |\bar\Delta_{\alpha \beta}|^2 = \Delta_0^2 $. For simplicity we will always choose the gauge with only $\bar \Delta_{12}=-\bar\Delta_{21} = \Delta_0$ with real $\Delta_0$ and $\bar\Delta_{23}=\bar\Delta_{13} =0$. Then the $\Delta$-sector of the matrix $\hat{M}$ in Eq. \ref{meq} is diagonal and we can just read off the spectral functions for the amplitude mode $\Delta^{r}_{12}$ and the phase mode $\Delta^{i}_{12}$ by inverting the corresponding matrix entries and taking their imaginary parts. For $\Delta_{13}$ and $\Delta_{23}$ the saddle points are at the origin. Therefore their real and imaginary parts fluctuate identically and have the same spectral function.  

Due to the constraint $\frac{1}{2} \sum_{\alpha,\beta} |\bar\Delta_{\alpha \beta}|^2 = \Delta_0^2 $ we expect soft modes for the 5 possible variations of $\Delta_{\alpha\beta}$ that leave $\Delta_0$ unchanged. Since only $\bar\Delta_{12} \not= 0$, these modes are the phase of $\Delta_{12}$, and real and imaginary fluctuations of $\Delta_{13}$ and $\Delta_{23}$. In addition to that there will be an amplitude mode of $\Delta_{12}$ at twice the gap frequency.

\begin{figure}
\includegraphics[width=.7\textwidth]{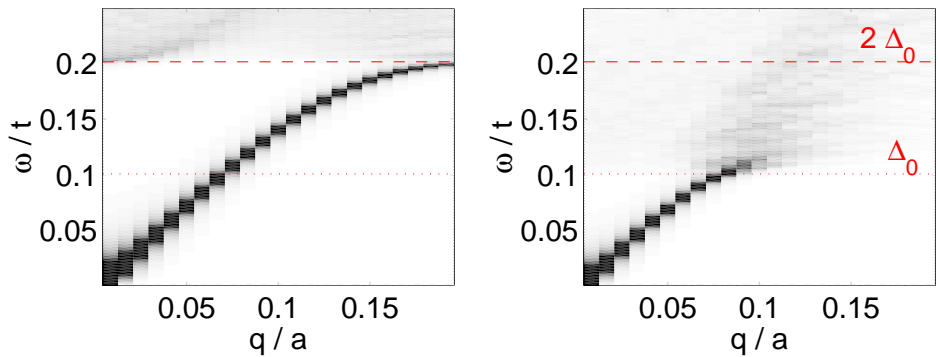}
\caption{Collective pairing modes in the SU(3)-symmetric case with onsite $s$-wave pairing $\bar\Delta_{12}=\Delta_0$, $\bar\Delta_{13}=\bar\Delta_{23}=0$ for 
$U_p=-4t$, filling $n = 0.556$ and $T=0.01t$. The steps in the data 
are due to the numerical evaluation for a finite number of wavevectors and frequencies.
Left plot: Spectral weight corresponding to fluctuations of $\Delta_{12}$. The
soft linear mode is the Anderson-Bogliubov phase mode, the weaker gapped feature at $2\Delta_0$ is due to amplitude fluctuations. 
Right plot:
Spectral weight of $\Delta_{13}$ or $\Delta_{23}$ fluctuations. 
Since the mean fields $\bar\Delta_{13}= \bar\Delta_{23} =0$, real and imaginary parts have the same fluctuation spectrum.   Landau-damping due to the gapless flavor 3 starts at $\omega =\Delta_0$. }
\label{CMU404040}
\end{figure} 

The results of the generalized RPA are shown in Fig. \ref{CMU404040}. In the left panel we find the phase mode of flavors 1 and 2 and the gapped amplitude mode. The Goldstone phase mode is gapless and start with a linear dispersion at small $q$. The velocity $c$ of the mode for our lattice model is very close to the continuum result $c=v_F/\sqrt{2}$ in two dimensions, where $v_F$ is the Fermi velocity along $\vec{q}$ ($c=v_F/\sqrt{3}$ in 3D)\cite{anbog}.
The amplitude mode has a gap of $2\Delta_0$ and is not sharp as it can decay into quasiparticle pairs above the gap edge. In this approximation the phase mode is undamped for $\omega < 2\Delta_0$. However we will show below that remaining density-density interactions for $U_\rho \not=0$ lead to a broadening of the phase mode. 
We also observe that the soft modes of $\Delta_{13} $ and $\Delta_{23}$ get strongly broadened above $\omega = \Delta_0$. Flavor 3 is ungapped and therefore these modes can excite quasiparticle pairs of flavor 1 and 3 (or 2 and 3) already above $\Delta_0$.

\subsection{Broken SU(3)-symmetry}
Here we briefly describe how two types of SU(3)--symmetry breaking in the 
Hubbard Hamiltonian affect the internal pairing modes.

The first case we consider is unequal values of the attraction between the flavors, for example  
$-U_{12} > -U_{13} > -U_{23}$. 
Then the mean-field solution chooses to pair the two flavors with the strongest mutual attraction. 
In the specific example this corresponds to $\bar\Delta_{12} \not= 0$,
while the other two pairing amplitudes are zero, $\bar\Delta_{13}=\bar\Delta_{23}=0$.

\begin{figure}
\includegraphics[width=.7\textwidth]{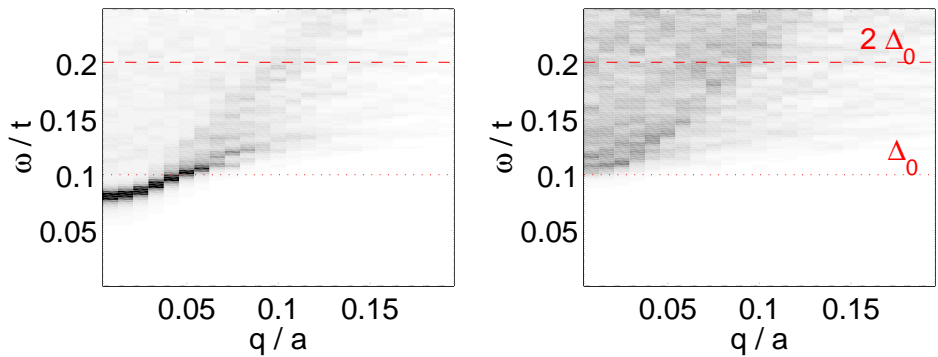}
\caption{Collective pairing modes in the SU(3)-broken case with onsite $s$-wave pairing and $\bar\Delta_{12}=\bar\Delta_0$, $\bar\Delta_{13}=\bar\Delta_{23}=0$, $U_{12}=4t$, $U_{13}=3.6t$,  and $U_{23}=2t$, $T=0.01t$. 
Left plot: Spectral weight corresponding to fluctuations of $\Delta_{13}$. The
dark feature is due to two gapped modes corresponding to fluctuations of the
real and imaginary part of $\Delta_{13}$ just below the particle-hole (PH) continuum. 
Right plot:
Spectral weight of $\Delta_{23}$ fluctuations. The attraction $U_{23}=2t$ in this channel is not strong enough to pull the modes below the PH continuum and the modes are almost wiped out by the PH-pairs.}
\label{CMU403620}
\end{figure} 
An analysis of the internal pairing modes gives the following picture.
The modes of this dominant $\Delta_{12}$-pairing channel remain unchanged. In particular the phase mode remains gapless. This is plausible as the unequal strength of the interactions does not lift the degeneracy of the ground state with respect to the phase of $\Delta_{12}$.
If the attraction in a subdominant channel is strong enough, the collective
oscillations of the pairing field in this channel get pulled below the
particle-hole continuum and hence get sharp. This is shown in Fig. \ref{CMU403620}, where the attraction in the 13-channel is strong enough to yield a sharp mode below $\Delta_0$, while the mode in the 23-channel with the least attractive interaction remains above the $\Delta_0$-threshold and is strongly damped. 
It may actually be interesting to see if the fluctuation energy (i.e. the frequency)  of these modes could be lowered by a modification of the mean-field state. Switching on a nonzero pairing amplitude in the subdominant channel will have an opposite effect as it pushes the modes further up. 

Next we consider the effects of a Zeeman splitting between the hyperfine levels. This corresponds 
to different chemical potentials or, equivalently, different densities of the three flavors. 
For simplicity let us assume a Zeeman shift $+E_Z$ for flavor 1 and $-E_Z$ for flavor 3, respectively, 
while the energy of flavor 2 remains unchanged.  For band filling of less than half filling, 
the pairing will select $\bar\Delta_{12}=\Delta_0$, as then flavors 1 and 2 have the larger densities 
of states.
As mentioned in Sec. II, the mean-field solution breaks down only when the band splitting $E_Z$ 
reaches the energy gap of the pairing $\Delta_0$.
Even in this case of SU(3) symmetry breaking, the degeneracy of the mean-field state with respect 
to the phase of $\Delta_{12}$ is not lifted. Therefore the phase mode of $\Delta_{12}$ remains 
a Goldstone mode. As the symmetry between the flavors is broken, the modes in the unpaired channels 
acquire gaps of the order of the Zeeman splitting.

\section{Density response}
Having studied the internal pairing modes of the SU(3) superfluid,
we now ask how these excitations couple to density oscillations. This
is important in order to find experimental signatures of the
collective excitations via Bragg scattering, as explained in 
section~\ref{RPA}. In fact, for the SU(2) case the appearance of 
the Anderson--Bogoliubov phase mode has been suggested as a 
promising way to detect the onset of fermionic superfluidity\cite{hofstetter}.

\subsection{Pure collective pairing fluctuations in the total density response} 
First, in order to separate the different effects, we study the
density response when the interaction is decoupled in the pairing
channel only, i.e.~we keep $U_\rho=0$ in Eq. \ref{decomp}. This means
that the only collective excitations stem from pairing fluctuations.
 Analyzing the imaginary part of $\chi_\rho (\vec{q},\omega)$ we find
 that only one of the six collective pairing modes show up in the
 density response. This mode is the phase mode of the nonzero pairing
 amplitude. This is not surprising. If the pairing amplitude in a
 certain channel is zero (say $\bar\Delta_{23}=0$), there is no superfluid
 density that could react to a phase fluctuation. Thus there is no
 quadratic coupling between density and phase.
The second observation is that the density signature of the phase mode
in the density response goes to zero for $q\to 0$. This can be
understood as follows. From the continuity equation we get $\delta
\rho \propto \frac{q^2}{\omega}\delta \phi$,
i.e. if $\omega \sim q$ the density change $\delta \rho $ induced by a
phase variation $\delta \phi$ vanishes like $q$. Consistent with that
we find that the density response of the phase mode becomes smaller
for $q \to 0$.
The amplitude mode of the paired flavors at $\omega= 2\Delta_0$ is not reflected in 
the density response, just as in the case of only two flavors\cite{varma}.

\begin{figure}
\includegraphics[width=.7\textwidth]{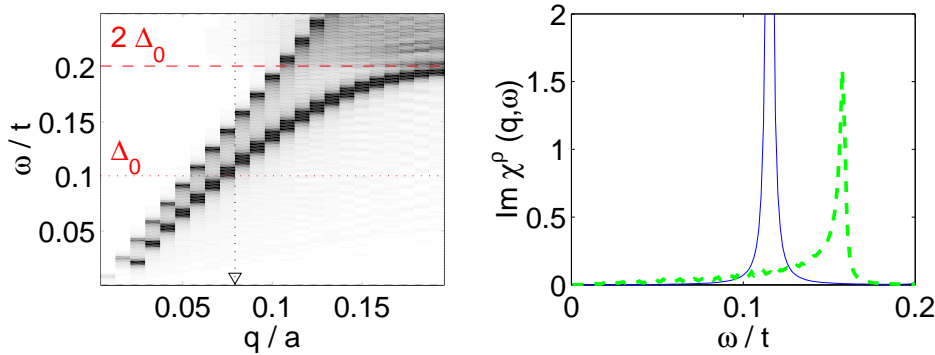}
\caption{Density response for the superfluid state at $T=0.01t$ and $\bar\Delta_{12}=\Delta_0$, 
$\bar\Delta_{13}=\bar\Delta_{23}=0$ 
in the case without remnant density-density interactions, $U_\rho=0$. Left plot: Im $\chi^\rho (\vec{q},\omega)$. The upper linear feature is the upper boundary of the particle-hole continuum of flavor 3, the lower feature is the density signature of the Anderson-Bogliubov phase mode. The amplitude mode and the modes involving $\Delta_{23}$ and $\Delta_{13}$ do not appear in the density response. 
Right plot: Cut through Im $\chi^\rho (\vec{q},\omega)$ for $q/a =0.08$ (see mark in left plot). The solid line is the density response of flavor 1 and 2, exhibiting the phase mode. The dashed line is the non-interacting particle-hole continuum of flavor 3. 
}
\label{chirhoU404040}
\end{figure} 
Thus for the case with only $\bar\Delta_{12}\not=0$ the density response consists of two features. The first is a sharp phase mode that -- in RPA -- is undamped below $2\Delta_0$, and the second is a particle-hole continuum of the 3rd unpaired flavor. Notice that in 2D the upper edge of the continuum is rather sharp.
This continuum feature distinguishes the density response of the 3-flavor case from that of the conventional 2-flavor superfluid. Below we will see how remaining density-density interactions affect this feature.

\subsection{Additional density-density interactions}
Next we want to analyze the effects of remaining density-density interactions.
So far we have decoupled the interaction in the pairing channel only, but as argued above collective density fluctuations can play a role as well. For that purpose we set $U_\rho=U_p=-4t$.

\begin{figure}
\includegraphics[width=.75\textwidth]{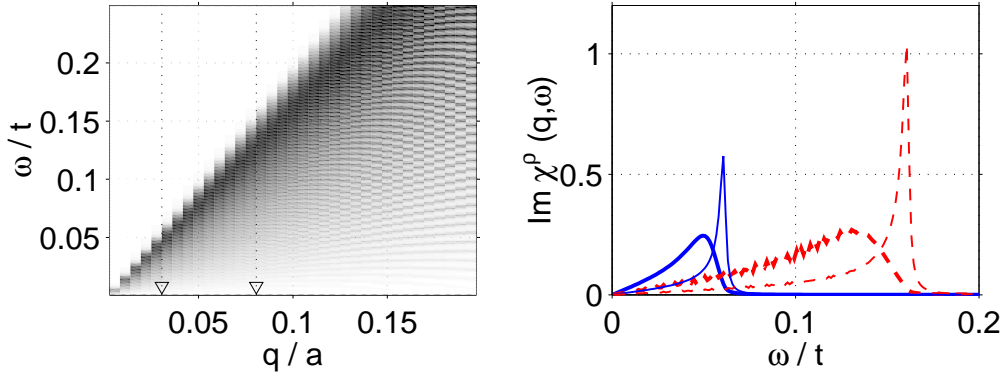}
\caption{Density response  for the normal state at $T=0.2t \sim 3T_c$ and 
$\bar\Delta_{12}= \bar\Delta_{13}=\bar\Delta_{23}=0$, 
for remnant density-density interactions $U_\rho=U_p=-4t$.
Left plot: Im $\chi^\rho (\vec{q},\omega)$ in the normal state when density-density interactions 
are included.  
Right plot: Cut through Im $\chi^\rho (\vec{q},\omega)$ for $q/a =0.03$ (solid lines) and $q/a=0.08$  (dashed lines, see marks in left plot). The thicker lines  correspond to the interacting case and the thin lines with the sharper peak show the non-interacting particle-hole continuum.
}
\label{RT123T21}
\end{figure} 
\begin{figure}
\includegraphics[width=.75\textwidth]{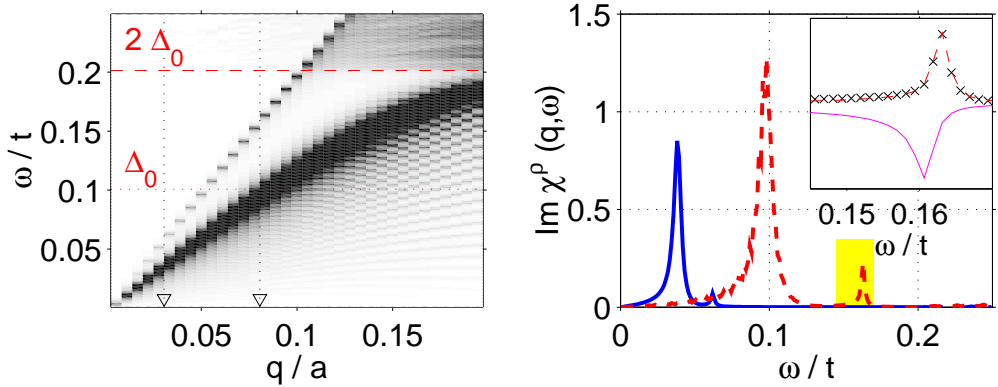}
\caption{Density response  for the superfluid state at $T=0.01t$ and $\bar\Delta_{12}=\Delta_0$, $\bar\Delta_{13}=\bar\Delta_{23}=0$, again for $U_\rho=-4t$. 
Left plot: Im $\chi^\rho (\vec{q},\omega)$. The upper linear feature is the signature of the flavor mode of $\rho_{123}$, the lower feature is the density signature of the Anderson-Bogliubov phase mode. The amplitude mode and the modes involving $\Delta_{23}$ and $\Delta_{13}$ do not show up in the density response. 
Right plot: Cut through Im $\chi^\rho (\vec{q},\omega)$ for $q/a =0.03$ (solid line) and $q/a=0.08$ (see marks in left plot). The inset is a zoom-up of the shaded region around $\omega/t=0.16$. The dashed line is the density response at $q/a=0.08$ and the crosses coinciding with it are the data points for the rescaled spectral function of the flavor mode $\rho_{123}$. 
For comparison, the other solid curve with the peak pointing downwards is proportional to the non-interacting particle-hole continuum.
}
\label{RT123T01}
\end{figure} 

In the normal state, we find two degenerate flavor modes corresponding to $\rho_{123}$ and $\rho_{12}$ which lie somewhat above the particle-hole continuum and are therefore undamped at low energies. However these modes do not show up in the density response. $\rho_t$ does not exhibit a sharp mode, as can be expected for attractive interactions. In this case, instead of creating a zero sound mode, the density-density interactions smear out the sharp upper edge of the bare particle-hole continuum, and the response of the total density exhibits only a broad peak.

In the paired state, the phase mode of $\Delta_{12}$ is the dominating feature in the total density response. In comparison to the results without density-density interactions we find that the coupling to the density fluctuations causes  slight damping of the mode. This can be seen by comparing the spectral functions in Figs. \ref{chirhoU404040} and \ref{RT123T01}. 
For our parameters and $U_\rho= U_p$ the phase mode is still clearly visible and distinguishes the density response of the paired state from the normal state result. In principle the temperature dependence of the width of this mode provides a way to detect the gapless third flavor. 
In addition to that, the density-density interactions renormalize the mode frequency to somewhat lower energies. In other words, in the paired state the gapping of flavors 1 and 2 sharpens up the zero sound mode that is pulled down by the interactions into the particle-hole continuum of flavor 3. 
Apart from the phase mode, there is a second feature in the density response that resembles the upper edge of the non-interacting particle-hole continuum.  
A closer comparison shows however that this second peak lies a bit above the free particle-hole continuum and is due to the flavor mode in the spectrum of $\rho_{123}$. This mode involves density variations $\propto n_1+n_2-2n_3$. Since for $\bar\Delta_{12} \not= 0$ flavor 3 has a distinct spectrum from flavors 1 and 2, this mode now couples to the total density $n_1+n_2+n_3$. Without the $\rho_{123}$ mode the density response would apart from the phase mode only exhibit a broad peak at the upper edge of the particle-hole continuum analogous to the normal state.

The data described above were obtained for equal partitioning of the interaction  with $U_p=U_\rho$. We have checked that other splittings of the local attraction with $U_\rho \not= U_p$ lead to the same qualitative results. For example reducing $U_\rho$ simply interpolates between the structure factor shown in Fig. \ref{chirhoU404040} and the data shown in Fig. \ref{RT123T01}. The damping of the phase mode increases with $|U_\rho|$. The additional flavor mode of $\rho_{123}$  merges continuously into the particle-hole continuum of flavor 3 when we let $U_\rho$ go to zero. Larger values of $|U_\rho|$ pushes the mode further above the continuum by increasing its linear dispersion with $q$. Therefore the damping of the Anderson-Bogoliubov mode as well as the $\rho_{123}$-feature near the upper edge of the particle-hole continuum of flavor 3 are robust results that do not depend on the ambiguity of the HS transformation in several channels. 

We  close the section by commenting on perturbations that break the SU(3)--symmetry  
of the Hamiltonian, like unequal interaction strengths between the flavors and Zeeman splitting. 
Their effect on the pairing modes was considered in the previous section. As the phase mode 
remains robust with respect to these perturbations and the single-particle excitations do not 
become altered qualitatively, we do not find significant changes in the density response 
as long as the mean-field state remains unchanged.

\section{Conclusions}
We have analyzed pairing and density fluctuations in the BCS paired
state of the attractive SU(3) Hubbard model.  This model can
potentially be realized by trapping ultracold fermionic atoms with 3
different hyperfine states (flavors) in an optical lattice. We have
studied the specific case of the two-dimensional square lattice, but
most results also hold for 3 spatial dimensions and even in the absence of an 
optical lattice (pure harmonic trapping). 

Even parity onsite BCS pairing in the SU(3) Hubbard model leads to a partially gapped quasiparticle spectrum. In the gauge where only $\bar\Delta_{12} \not= 0$, two flavors get gapped and the third flavor retains its full Fermi surface. As the pairing amplitude $\Delta_{\alpha \beta}$ transforms non-trivially under SU(3)$\otimes$ U(1), in addition to the usual Anderson-Bogoliubov phase mode there are 4 other Goldstone modes. 
However only the phase mode corresponding to the two gapped quasiparticle branches couples to the density of the fermions.  Hence only one of the 5 Goldstone modes is visible in the dynamic structure factor $S(\vec{q},\omega)$. This should be observable e.g. in Bragg scattering experiments. 
For the case of two flavors with local attractions, it has been shown that the emergence 
of the phase mode provides a clear signature of the pairing. This occurs because 
in the normal state the zero sound mode gets wiped out by low--lying particle--hole excitations 
which are gapped in the paired state.
If a third flavor is present this mode remains broadened at all temperatures 
due to the damping by the ungapped quasiparticle branch. However, it is clearly visible in 
the Bragg response. 

In addition to the phase mode we find another marked feature in the density response which arises 
due to a flavor mode where the densities of the two gapped flavors oscillate out of phase with 
the density of the ungapped flavor. This mode has a linear dispersion and lies somewhat above 
the particle-hole continuum of the gapless branch of the fermionic excitation spectrum. 
The separation from the continuum depends on the strength of the remnant density-density interaction 
in the paired state.
Together with the visible damping of the phase mode at sub-gap frequencies at all temperatures, 
this flavor mode feature should render a clear signature of the ungapped part of the 
quasiparticle spectrum and thus of the presence of three degenerate flavors in the system.

We have furthermore investigated the effect of perturbations that break the SU(3) symmetry 
of the Hamiltonian. For example a Zeeman splitting, leading to different chemical 
potentials of the three flavors, only destroys the pairing when it 
becomes comparable to the energy gap due to pairing, $\Delta_0$. 
The density response remains qualitatively the same as long as the mean-field state is unchanged. 
This gives confidence that the described phenomena should remain robust over a wider parameter range.

Acknowledgments: We thank W.~Ketterle, P.A.~Lee, M.~Salmhofer, R.~Zeyher, P.~Zoller and M.~Zwierlein for 
useful discussions. W.H. was supported by a Pappalardo fellowship.

\end{document}